\documentclass[12pt]{iopart}
\usepackage{iopams}
\usepackage{cite}
\usepackage{graphicx}
\usepackage{eufrak}

\begin{document}

\title[Regularization of the big bang singularity with random perturbations]{Regularization of the big bang singularity with random perturbations}

\author{Edward Belbruno$^1$, BingKan Xue$^2$}

\address{$^1$ Department of Astrophysical Sciences, Princeton University, Princeton, NJ 08544, USA \footnote{ \& Department of Mathematics,  Yeshiva University,  New York, NY 10033 }}
\address{$^2$ School of Natural Sciences, Institute for Advanced Study, Princeton, NJ 08540, USA}

\eads{\mailto{belbruno@princeton.edu}, \mailto{bkxue@ias.edu}}

\begin{abstract}

We show how to regularize the big bang singularity in the presence of random perturbations modeled by Brownian motion using stochastic methods. We prove that the physical variables in a contracting universe dominated by a scalar field can be continuously and uniquely extended through the big bang as a function of time to an expanding universe only for a discrete set of values of the equation of state satisfying special co-prime number conditions. This result significantly generalizes a previous result \cite{Xue:2014} that did not model random perturbations.  This result implies that the extension from a contracting to an expanding universe for the discrete set of co-prime equation of state is robust, which is a surprising result.  Implications for a purely expanding universe are discussed, such as a non-smooth, randomly varying scale factor near the big bang.

\end{abstract}

\submitto{\CQG}

\maketitle

\section{Introduction} \label{sec:intro}

We consider the classical Friedmann equations that describe the evolution of the universe. We consider the scenario where the universe transitions from contraction to expansion through the big bang. This evolution is described by dynamics of the following quantities:  $a, H, w, \Omega$, which are the scale factor, Hubble parameter, equation of state, and relative density of energy components (representing matter, radiation, cosmological constant, spatial curvature and anisotropy), respectively. $w$ is defined by a scalar field, $\phi$. These quantities vary as a function of time, $t$, where the big bang occurs at $t=0$. It is assumed that the universe contracts for $t < 0$ and expands for $t > 0$ as in a bouncing universe model. The Friedmann equations are not defined at the big bang. 

We show how to regularize the big bang singularity in the presence of random perturbations that are modeled using Brownian motion in a full neighborhood of the big bang singularity. In this modeling the random perturbations occur continuously from moment to moment up to and including the big bang itself. This is done by using a stochastic model for the Friedmann equations together with a stochastic version of the stable manifold theorem due to Mohammed and Scheutzow \cite{Mohammed:1999}. This allows the study of solutions of the Friedmann equation, $a(t), H(t), w(t), \Omega(t)$,  as they pass through the big bang from $t < 0$ to $ t > 0$ where their dynamics can be understood.
It is noted that adding random perturbations to the Friedmann equations can model effects that are known as well as unknown. This is analogous to a technique done in celestial mechanics where  motion of the planets is modeled using an interpolated planetary ephemeris, which continuously models all the empirical random perturbations on the planetary orbits, even ones that are not known \cite{Belbruno:2004}.

A key result of this paper is to prove that this extension of solutions through $t=0$ is only possible, and unique,  if the value of the equation of state at the big bang, $w_c = w(0)$, takes on a value of a special discrete set, $\mathbb{P}_w$, of co-prime integers, where it is assumed that $w_c > 1$ \cite{Garfinkle:2008ei}, \cite{Erickson:2003zm}. This result is stated as Theorem 2 in Section \ref{sec:4}. This result is a significant generalization of \cite{Xue:2014}, where random perturbations were not modeled. In that case, the same condition that $w_c$ had to take on a value from  $\mathbb{P}_w$ was obtained; however, that result was limited since random perturbations were not modeled.  That opened up the possibility that if random perturbations were included, then the co-prime condition would no longer be well defined due to the random fluctuations given by the Brownian motion.  

It is surprising that this co-prime condition is still required when random perturbations are included since it represents a fine tuning of the solutions at the big bang which is preserved under random fluctuations. This suggests that the co-prime condition may be physically significant and not only due to the mathematics. This paper provides a nontrivial example of the fact that even in the presence of random perturbations, the evolution of the universe can be continued through the big bang for $a=0$. 

Although the random perturbations added to the Friedmann equations do not change the metric we are using, defined in Section \ref{sec:1}, they do give considerable complexity to the dynamics of the scale factor $a(t)$, which can be viewed as analogous to perturbing the metric. This is discussed in Section \ref{sec:5}.

An interesting result is obtained on the behavior of the scale factor $a(t)$  near the big bang. It is proven that $a(t)$ need not be smoothly monotone decreasing or increasing sufficiently near the big bang. In particular, if we just consider the behavior of our solutions in the case of a purely expanding universe without regards of how it initiated, $a(t)$ can have random, non-smooth, jagged variations given by the random perturbations. This result is described at the end of Section \ref{sec:4}, and expressed analytically in Equation \ref{eq:AofTStochastic}. We discuss how it may be possible to detect this experimentally in Section \ref{sec:5}.

The random perturbations we consider are defined by a general Brownian motion, which can be arbitrarily specified by a set of functions of the random values. They exist up to and including the big bang. In Section \ref{sec:5},  we discuss how they may be used to model quantum fluctuations.

The method of obtaining these results is to first regularize the Friedmann equations by a transformation of coordinates and time so that, in the new variables, the transformed differential equations are defined at the big bang. This transformation reduces the big bang to a hyperbolic fixed point for the flow of the solutions. This approach was used in \cite{Xue:2014}. In the regularized coordinates, it can be shown that solutions are defined from $t \leq 0$ to $t \geq 0$ if and only if $w_c \in \mathbb{P}_w$. However, in the current situation, this approach is significantly more complex. To include random perturbations to the variables, it is necessary to view the coordinates as being random variables in a probability space. Then, the random perturbations can be introduced in a well defined manner by writing the Friedmann equations in a special stochastic form, where the random perturbations are defined by the Brownian motion written in a particular from. A key part of the analysis is to show that the flow of the stochastic differential equations are structurally stable in a neighborhood of the hyperbolic fixed point. To show this,  a stochastic version of the stable manifold theorem \cite{Mohammed:1999} is used which places conditions on the form of the functions defining the random perturbations.

In Section \ref{sec:1}, we describe the assumptions for the modeling of the classical Friedmann equations. In Section \ref{sec:2}, the previous results from \cite{Xue:2014} are summarized. Stochastic modeling and definitions are given in Section \ref{sec:3}, where the transformation of the Friedmann equations to stochastic It\^{o} equations is described. The stochastic stable manifold theorem is described in Section \ref{sec:4} and applied to the differential equations to obtain the main result on the extension of solutions through the big bang. In Section \ref{sec:n}, the main results are numerically verified.   In Section \ref{sec:5}, the results are discussed.  Finally, Section \ref{sec:6} is the Conclusion.

Some results and conclusions are distinguished by italicized subheadings.

\section{Modeling and Assumptions} \label{sec:1} 

In this section, we recall from \cite{Xue:2014} the differential equations for a set of variables that describe the evolution of the universe. The Friedmann equations lead to a differential equation for the Hubble parameter $H$, or its reciprocal, $Q \equiv 1/H$. The equation of motion for a scalar field $\phi$ determines its time varying equation of state $w$. With additional energy components besides the scalar field, we introduce the relative energy density $\Omega_m$ for each component.

We assume a homogeneous, flat, and isotropic universe with the metric
\begin{equation}
ds^2 = - dt^2 + a(t)^2 |d{\bf x}|^2 ,
\label{eq:metric}
\end{equation}
where $t$ is the proper time and $\mathbf{x} = (x^1, x^2, x^3)$ are the spatial coordinates. Here $a(t)$ is the scale factor of the universe, and the Hubble parameter $H$ is given by $H \equiv \dot{a} / a$, where the dot $\dot{}$ denotes the derivative with respect to time $t$. $H$ is negative during cosmic contraction, where $ H \rightarrow -\infty$ as $ t \rightarrow 0^-$ and positive after the universe transitions to expansion, where $H \rightarrow \infty$ as $ t \rightarrow 0^+$. The big bang (or ``big bounce'') is at $a = 0$, which is chosen to correspond to the time $t = 0$.

We assume that the scalar field $\phi$ has the Lagrangian
\begin{equation}
\mathcal{L} = \sqrt{-g} \Big[ -\case{1}{2} (\partial_t \phi)^2 - V(\phi) \Big] ,
\end{equation}
where the potential $V(\phi)$ is an exponential function $V(\phi) = - V_0 \, e^{- c \, \phi}$; it is assumed that $V_0 > 0$ and $c$ is a constant, whose range is determined in the following text. In the homogeneous case, the energy density and pressure of the scalar field are
\begin{equation}
\rho_\phi = \case{1}{2} \, \dot{\phi}^2 + V(\phi), \quad p_\phi = \case{1}{2} \, \dot{\phi}^2 - V(\phi).
\label{eq:rhoP}
\end{equation}
The equation of state parameter $w$ is given by
\begin{equation}
w = {p_\phi} / {\rho_\phi} .
\label{eq:w}
\end{equation}
For our study, we focus on the case $ w > 1$ and $\rho_{\phi} > 0$; hence $V(\phi) < 0$. \footnote{Such a negative exponential potential is typical of {\it ekpyrotic} models \cite{Steinhardt:2001st} \cite{Lehners:2011kr}.}

In addition to the energy component $\phi = \phi(t)$, with associated equation of state, $w=w(t)$, we assume other energy components with constant  equations of state $w_m$, $m=1,2,3,4,5$, where $w_m = 0$, $\frac{1}{3}$, $-1$, $-\frac{1}{3}$, or $1$, if the additional energy components represent matter, radiation, cosmological constant, spatial curvature, or anisotropy, respectively.  The energy density of each component is given by $\rho_m$. We define fractional energy density parameters,
$\Omega_m = \Omega_m(t)$,
\begin{equation}
\Omega_m \equiv \rho_m / \rho_{\rm tot} , \quad \rho_{\rm tot} = \rho_\phi + \sum_m \rho_m .
\end{equation}
It is noted that $w_m \leq 1$. \footnote{This requirement also guarantees that the speed of sound for this component is non-superluminal; indeed, for a constant $w_m \leq 1$, the speed of sound for this component is $c_m^2 \equiv \delta p_m/\delta \rho_m = w_m \leq 1$. Incidently, for the(canonical) scalar field, although $w(t) > 1$, the speed of sound is always $c_s^2 =1$ (see, e.g., \cite{Mukhanov:2005}).}

As is shown in \cite{Xue:2014}, in a contracting universe, where $H<0$,  $w(t) \rightarrow w_c$ as $t \rightarrow 0^-$, where
\begin{equation}
w_c \equiv {c^2\over{3}} -1.
\end{equation}
It can be seen that for $c > \sqrt{6}$,  $w_c > 1$. It is shown in \cite{Xue:2014} that $w = w_c$ is a fixed point attractor in a contracting universe.

We will consider in this paper one energy component in addition to $\phi$, which we label $\Omega_1$, with constant equation of state $w_1$. We set $W = w - w_c$. As $t \rightarrow 0^-$, this implies $H \rightarrow -\infty, W \rightarrow 0, \Omega_1 \rightarrow 0$. For simplicity of notation, we set $\Omega_1 \equiv \Omega$. It is shown in \cite{Xue:2014} that we obtain the system of differential equations,
\medskip

\noindent
\numparts
\begin{eqnarray}
\hspace{-0.5in} \dot{H} = -\case{3}{2}H^2 \Big[ (W + w_c + 1) - (W + w_c - w_1) \Omega \Big] , \label{eq:DiffEqusMoreGeneralH} \\*[4pt]
\hspace{-0.5in} \dot{W} = \case{3 H\sqrt{W + w_c + 1}}{\sqrt{W + w_c + 1} + \sqrt{(w_c + 1) (1 - \Omega_1)}} \, (W + w_c - 1) \big( W + (1 + w_c) \Omega \big) , \label{eq:DiffEqusMoreGeneralW} \\*[4pt]
\hspace{-0.5in} \dot{\Omega}= 3H(W + w_c - w_1) \Omega (1 - \Omega) . \label{eq:DiffEqusMoreGeneralOmega}
\end{eqnarray}
\endnumparts

\medskip

\noindent
\textit{Remark.} \; It is noted that the above case can be generalized to include more energy components with constant equations of state $w_m$, $m = 1, 2, \cdots$. The generalized differential equations for $H$, $W$, and $\Omega_m$ are given by
\numparts
\begin{eqnarray}
\hspace{-1in} \dot{H} = -\case{3}{2} H^2 \bigg[ (W + w_c + 1) - \sum_m (W + w_c - w_m) \Omega_m \bigg] , \label{eq:DiffEqusMostGeneralH} \\*[4pt]
\hspace{-1in} \dot{W} = \case{3H \sqrt{W + w_c + 1} }{\sqrt{W + w_c + 1} + \sqrt{(w_c + 1) (1 - \sum_m \Omega_m)}} \, (W + w_c - 1) \Big( W + (1 + w_c) \sum_m \Omega_m \Big) , \label{eq:DiffEqusMostGeneralW} \\*[4pt]
\hspace{-1in} \dot{\Omega}_m = 3H \Omega_m \bigg[ (W + w_c - w_m) (1 - \Omega_m) - \sum_{n \neq m} (W + w_c - w_n) \Omega_n \bigg] , \quad m = 1, 2, \cdots \label{eq:DifEqusMostGeneralOmega}
\end{eqnarray}
\endnumparts
where the summation is over the additional energy components besides the scalar field.

\medskip

\section{Classical Regularization, Previous Results} \label{sec:2}

In this section we summarize previous results in \cite{Xue:2014} on the regularization of System \ref{eq:DiffEqusMoreGeneralH}, \ref{eq:DiffEqusMoreGeneralW}, \ref{eq:DiffEqusMoreGeneralOmega} in a neighborhood of the big bang, and  consider the solutions $a(t)$, $Q(t) = H(t)^{-1}$, $W(t)$, and $\Omega(t)$ for $t < 0$, which tend to $a = 0$, $Q = 0$, $W = 0$, and $\Omega = 0$ when $t \to 0^-$. We determine necessary and sufficient conditions for these solutions to have a well defined unique extension to $t \geq 0$.

We use regularization methods that are traditionally used in classical and celestial mechanics. There are different types of regularizations \cite{Belbruno:2004}, \cite{McGehee:1981} which all require a change of variables as well as a time transformation. The strongest type of regularization, {\it global regularization}, transforms the singularity into a regular point at which the transformed differential equations become well defined, and real analytic, in a full neighborbood, of the singularity. The transformed solutions can be extended through the singularity in a real analytic manner, both with respect to time and initial conditions. One such example is the collision of two point masses in the Newtonian two-body problem, which can be globally regularized to describe a smooth bounce by using the Levi-Civita transformation. Another type of regularization, {\it branch regularization} reduces the singularity to a well defined finite point. This point can also be a fixed point, although this is not necessary.  Accordingly, the trajectory of an individual solution flowing to the finite point can be uniquely matched to the trajectory of another solution emerging from the same point. For such a regularization, the solutions in the original variables and time can be extended through the singularity in a continuous manner with respect to time. There are also singularities that cannot be regularized at all. In such cases, not a single solution can be extended through the singularity. This occurs, for example, in a triple collision in the Newtonian three-body problem.

System \ref{eq:DiffEqusMoreGeneralH}, \ref{eq:DiffEqusMoreGeneralW}, \ref{eq:DiffEqusMoreGeneralOmega} can be branch regularized. This means that there exists a transformation of the coordinates and time such that in the new coordinates and time, the big bang state is well defined at a finite time and there exists initial conditions for solutions in the contracting universe for $t<0$ that can be continuously connected in a unique manner to solutions for $t \geq 0$. This turns out to be true for a special set of values of $w_c$. The transformation of coordinates that is used maps the big bang at $|H| = \infty, W =0, \Omega =0$ into a hyperbolic fixed point for the transformed system of differential equations.

The main result proven in \cite{Xue:2014} is,
\medskip\medskip

\noindent
\textbf{Theorem~1.} \; The solutions $H(t), W(t), \Omega(t)$ of the dynamical system \ref{eq:DiffEqusMoreGeneralH}, \ref{eq:DiffEqusMoreGeneralW}, \ref{eq:DiffEqusMoreGeneralOmega}, as well as $a(t)$, are branch regularizable at the singularity $t = 0$ if and only if the value of $w_c$ belongs to a discrete set $\mathbb{P}_w$ given by
\begin{equation}
\mathbb{P}_w = \Big\{ w_c =  \case{2 q}{3 p} - 1 \; \Big| \; p, q \in \mathbb{Z}^+ , \; p < q, \; p \perp q, \; q \mbox{ odd} \Big\} ,
\end{equation}
where  $p \perp q$ means that the integers are co-prime.
\medskip\medskip\medskip

\begin{figure}[h!]
\centering
	\includegraphics[width=0.45\textwidth, clip, keepaspectratio]{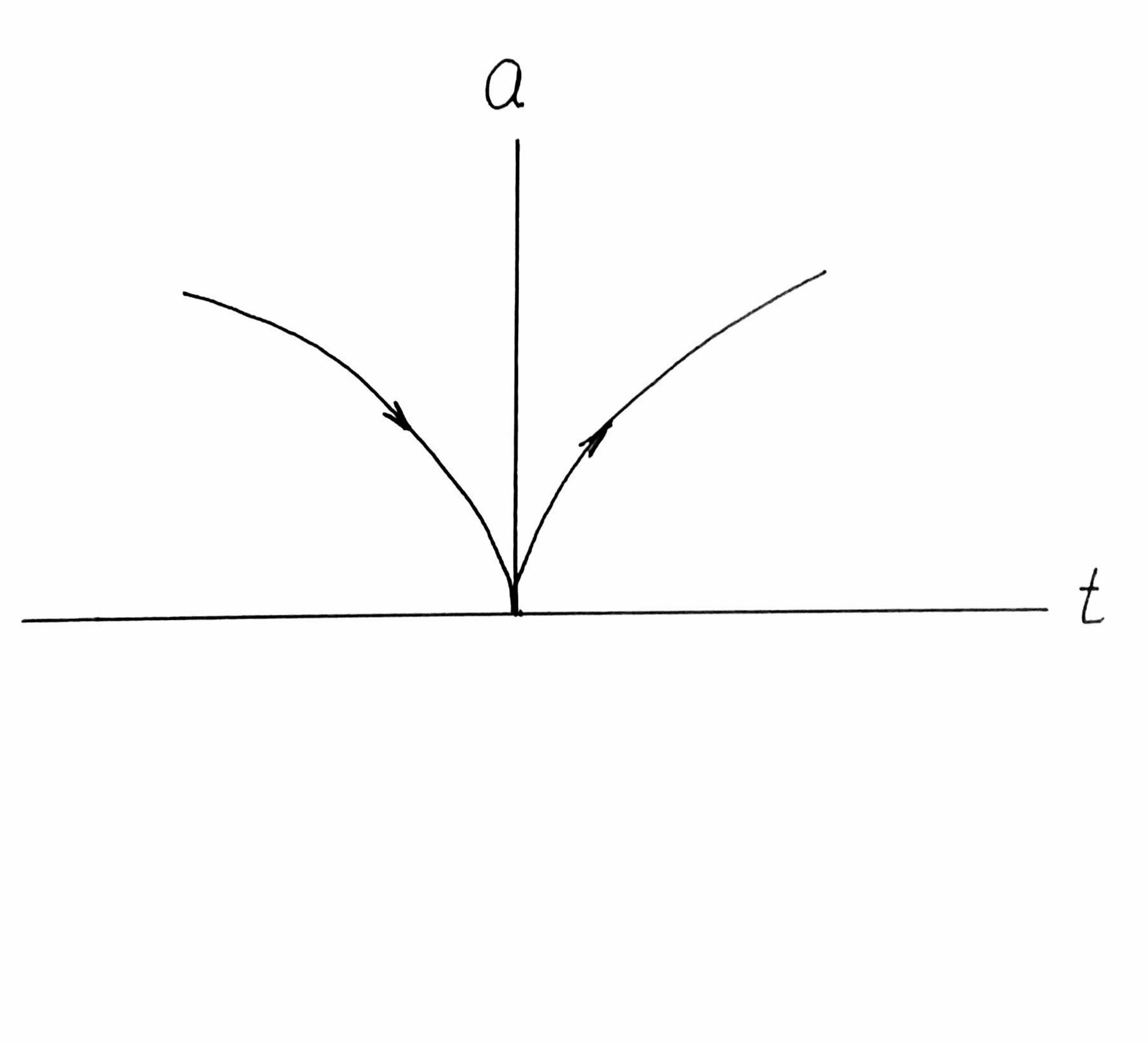}
\caption{Behavior of $a(t)$ in a neighborhood of $t=0$ where $w = w_c \in \mathbb{P}_w.$ }
\label{fig:1}
\end{figure}

Dynamically, if we view this theorem in terms of the variable $a(t)$, then for $t < 0$ in the contracting universe, $ a(t) \rightarrow 0$ as $ t \rightarrow 0^-$. Also, $w(t) \rightarrow w_c$. If we assume that $w_c \in \mathbb{P}_w$, then at $t=0$, $a = 0$ and $w = w_c$.  $a(t)$ increases from $0$ for $t >0$ in the expanding universe.  $\dot{a} \rightarrow  \infty$ as $|t| \rightarrow 0$. The graph of $a(t)$ will show a cusp at $t=0$, but the curve for $a(t)$ is continuously connected at $t=0$. See Figure ~\ref{fig:1}.   \footnote{Note that the variable $H(t)$ is not defined at $t = 0$; however, its solution for $t < 0$ has a unique branch extension to $t > 0$.}
\medskip\medskip\medskip

It is instructive to outline a sketch of how the theorem is proven since we will refer back to it in the translation of this problem to a stochastic system in Section \ref{sec:3}.

The regularization transformation, in part, is given by,
\begin{eqnarray}
H = Q^{-1} , \label{eq:HMap} \\
 {{dt}\over{d\tau}} = -Q,   \label{eq:TimeMap}
\end{eqnarray}
where, $Q < 0$, $t < 0$.  The time transformation, $t \rightarrow \tau$ can equivalently be written as 
\begin{equation}
\tau = - \ln a + \hbox{const} .
\label{eq:NTsolution}
\end{equation}
It is noted that $\tau > 0$ for $t$ sufficiently near $0$.
The System  \ref{eq:DiffEqusMoreGeneralH}, \ref{eq:DiffEqusMoreGeneralW}, \ref{eq:DiffEqusMoreGeneralOmega}  transforms into,
\numparts
\begin{eqnarray}
\hspace{-0.5in} \frac{dQ}{d\tau} = - \case{3}{2} \Big[ (W + w_c + 1) - (W + w_c - w_1) \Omega \Big] Q , \label{eq:dQdN} \\*[4pt]
\hspace{-0.5in} \frac{dW}{d\tau} = \case{- 3 \sqrt{W + w_c + 1}}{\sqrt{W + w_c + 1} + \sqrt{(w_c + 1) (1 - \Omega)}} \, (W + w_c - 1) \big( W + (1 + w_c) \Omega \big) , \label{eq:dwdN} \\*[4pt]
\hspace{-0.5in} \frac{d\Omega}{d\tau} = - 3 (W + w_c - w_1) \Omega (1 - \Omega) . \label{eq:dOmegadN}
\end{eqnarray}
\endnumparts
\medskip\medskip\medskip

We set $\mathbf{X} = (Q,W,\Omega)$ and let $\mathbf{X}(\tau, \mathbf{X_0})$ represent a solution of System \ref{eq:dQdN},\ref{eq:dwdN}, \ref{eq:dOmegadN} with initial condition $\mathbf{X_0}$, which we write as $\mathbf{X}(\tau)$ for short, where $\mathbf{X}_0 $ is near zero in norm.  This initial condition is for a value of $\tau_0 > 0$ corresponding to $t _0 < 0$ in the contracting universe, where $|t_0|$ is small, or equivalently, $\tau > 0$  is sufficiently large as follows from (\ref{eq:NTsolution}) since $a(t_0)$ is near zero, or equivalently.

The big bang singularity is given by $\mathbf{X} = (0,0,0) \equiv \mathbf{0}$, at $t =0$.  As is described in \cite{Xue:2014}, $\mathbf{X} = \mathbf{0}$ is a fixed point, or equivalently a stationary point, for the flow(solutions) of System \ref{eq:dQdN},\ref{eq:dwdN}, \ref{eq:dOmegadN}, and $t \rightarrow 0^-$ is equivalent to $\tau \rightarrow \infty$. Also, if this system is linearized at 
$\mathbf{X} = \mathbf{0}$, then the fixed point is locally hyperbolic with purely negative real eigenvalues. More precisely, the system can be written near $\mathbf{X} = \mathbf{0}$ as
\begin{equation}
\frac{d \, \mathbf{X}}{d\tau} = \mathbf{A} \mathbf{X} + \mathbf{\Delta}(\mathbf{X}) ,
\label{eq:NonlinearSystem}
\end{equation}
where 
\begin{equation}
\mathbf{A} = \left( \begin{array}{ccc}
-\case{3}{2} (w_c + 1) & 0 & 0 \\[4pt]
0 & -\case{3}{2} (w_c - 1) & -\case{3}{2} (w^2_c - 1) \\[4pt]
0 & 0 & -3 (w_c - w_1) \end{array} \right) ,
\label{eq:MatrixA}
\end{equation}
and the vector field $\mathbf{\Delta}(\mathbf{X})$ given by
\begin{equation}
\hspace{-1in} \mathbf{\Delta}(\mathbf{X}) = \left( \begin{array}{c} -\case{3}{2} Q \big[ W - (w_c + W - w_1) \Omega \big] \\[4pt]
-\frac{3 \big( 1 + 3 w_c + 2 W + 2 \sqrt{(1 + w_c) (1 + w_c + W) (1 - \Omega)} \big) W + 3 (w_c^2 - 1) \Omega}{2 \big( \sqrt{(1 + w_c) (1 - \Omega)} + \sqrt{1 + w_c + W} \big)^2} \, \big( W + (1 + w_c) \Omega \big) \\[4pt]
-3 W \Omega + 3 (w_c - w_1) \Omega^2 + 3 W \Omega^2 \end{array} \right) ,
\label{eq:FCase3}
\end{equation}
from which it is clear that $\mathbf{\Delta} = \mathcal{O}(|\mathbf{X}|^2)$. The eigenvalues of $\mathbf{A}$ are given by $\lambda_1 = -\frac{3}{2}(w_c-1)$,  $\lambda_2 = -\frac{3}{2}(w_c+1)$, $\lambda_3 = -3 (w_c - w_1)$. They are all negative since, by assumption, $w_c > 1 \geq w_1$.  It is seen that for $\mathbf{X}$ near $\mathbf{0}$, $\Delta(\mathbf{X})$ is a real analytic function of $\mathbf{X}$.  It is seen that when $\mathbf{X} =\mathbf{0}$, $\mathbf{\Delta}(\mathbf{0}) = \mathbf{0}$ and ${  {\partial \mathbf{\Delta}}\over{\partial \mathbf{X}}  }(\mathbf{0}) = \mathbf{0}$ ($3 \times 3$ matrix of 0).

In the linearized system where $\mathbf{\Delta}(\mathbf{X}) =0$, all the solutions exponentially flow towards $\mathbf{X} = \mathbf{0}$. More precisely, a basis of these solutions are given by the eigenvectors of the linear system,  $(\bar{Q}(\tau),0,0)$, $(0, \bar{W}(\tau),0)$, $(0,0, \bar{\Omega}(\tau))$, where
\begin{equation}
\bar{Q}(\tau) \propto - e^{-\frac{3}{2}(w_c+1)\tau} , \quad \bar{W}(\tau) \propto e^{-\frac{3}{2}(w_c-1)\tau} , \quad \bar{\Omega}(\tau) \propto e^{-3(w_c-w_1)\tau} .
\label{eq:Eigenvectors}
\end{equation}
These basis solutions are called hyperbolic invariant manifolds of the linearized system. (In the same way, we can show that for $t >0$, the solutions flow away from the fixed point, where the basis solutions  in that case are given (\ref{eq:Eigenvectors}) with plus signs in the exponents, and a plus sign in front of the exponential for $\bar{Q}(\tau)$ since  $\bar{Q}(\tau) > 0$. This follows since in this case the eigenvalues are given by $\lambda_1 = \frac{3}{2}(w_c-1)$,  $\lambda_2 = \frac{3}{2}(w_c+1)$, $\lambda_3 = 3 (w_c - w_1)$.  ) The eigenvectors of the linearized system with negative eigenvalues span a  subspace in which all points flow towards the fixed point as $\tau \rightarrow \infty$. This is called a {\it stable} subspace since all the points flow towards the fixed  point. This subspace has a manifold structure and is referred to as a \emph{stable manifold} and is 3-dimensional. It is referred to as a 3-dimensional hyperbolic invariant manifold, labeled $\mathbf{\bar{W}}^s$.  Similarly, if we replace $\tau$ by $\tilde{\tau} = -\tau$ in (\ref{eq:Eigenvectors}), then as $\tilde{\tau} \rightarrow \infty$, all the points flow away from the stationary point since, as mentioned, (\ref{eq:Eigenvectors}) has positive exponents where the eigenvalues are positive and also $\bar{Q} > 0$ . This yields a 3-dimensional set of {\it unstable} points called a hyperbolic unstable manifold, labeled $\mathbf{\bar{W}^u}$. 

\begin{figure}[h!]
\centering
	\includegraphics[width=0.45\textwidth, clip, keepaspectratio]{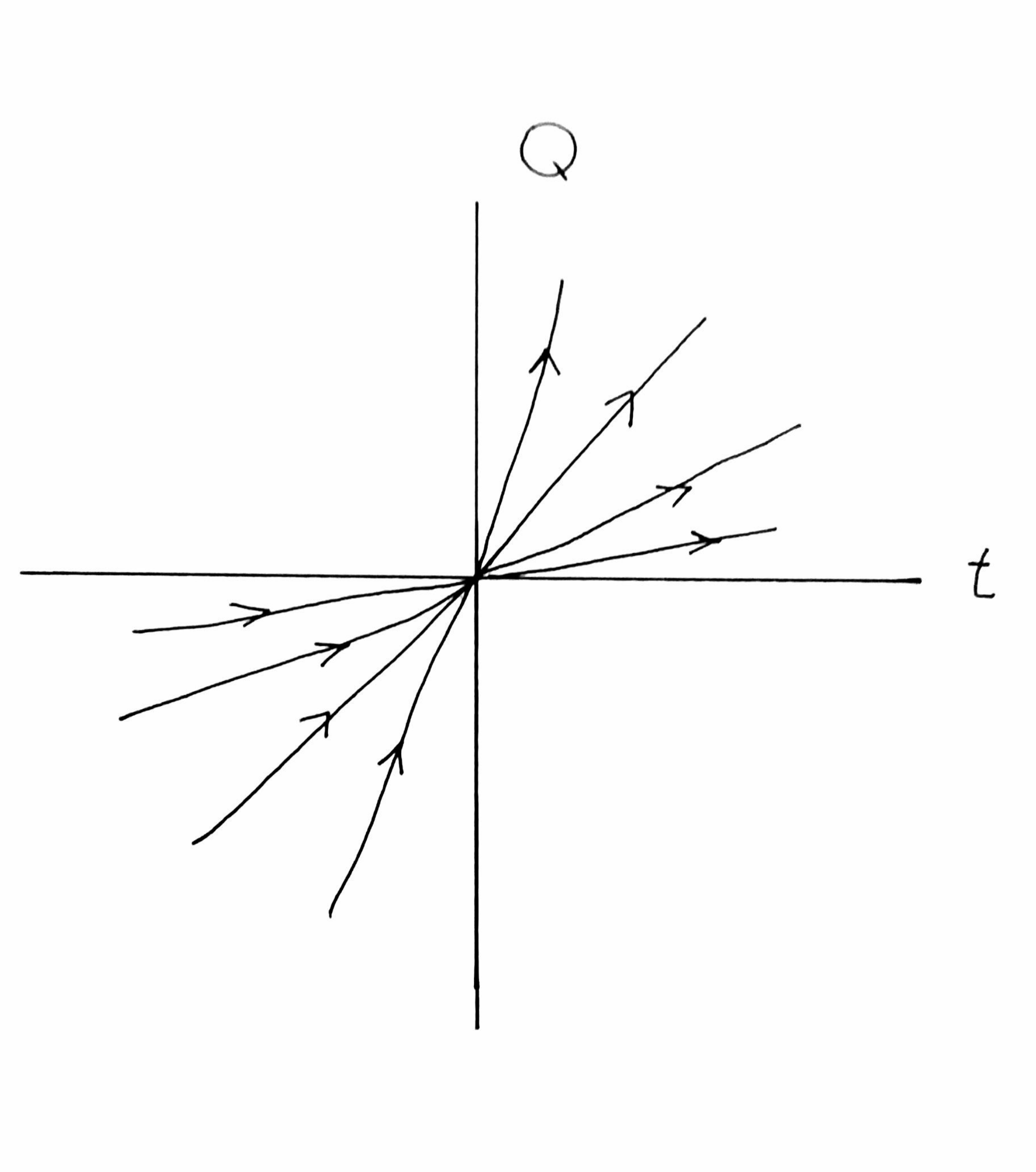}
\caption{Approaching the big bang as $\tau \rightarrow \infty$ ($t \rightarrow 0^-$) in contracting phase and expansion from big bang as $\tilde{\tau} \rightarrow -\infty$ ($t \rightarrow 0^+$).}
\label{fig:2}
\end{figure}

The {\it Stable Manifold Theorem}  describes what dynamically happens in a small neighborhood of $\mathbf{X} = \mathbf{0}$ when  $\mathbf{\Delta} \neq \mathbf{0}$, under the conditions $\mathbf{\Delta}(\mathbf{0}) = \mathbf{0}$ and ${  {\partial \mathbf{\Delta}}\over{\partial \mathbf{X}}  }(\mathbf{0}) = \mathbf{0}$.   When we consider $t <0$ , we would like to know what happens to the flow of (\ref{eq:NonlinearSystem}) near  $\mathbf{X} = \mathbf{0}$.  This theorem says that the basis solutions (\ref{eq:Eigenvectors}) of the linear system($\Delta = \mathbf{0}$) change by very little for $\Delta \neq \mathbf {0}$ provided $|\mathbf{X}|$ is sufficiently small. In other words, $\mathbf{\bar{W}}^s$ changes by very little, and similarly,  $\mathbf{\bar{W}^u}$ changes by very little, resulting in sets  $\mathbf{W}^s$ ,  $\mathbf{W}^u$.  This implies that the flow of  the nonlinear system (\ref{eq:NonlinearSystem}) changes by very little from the flow of the linear system for  $|\mathbf{X}|$ sufficiently small, i.e. sufficiently near the big bang.  Under assumptions on smoothness of $\Delta(\mathbf{X})$,  the flow of the solutions with respect to time will also  be real analytic, for $t < 0$ or $t >0$ in the nonlinear system. The manifolds  $\mathbf{W}^s$, $\mathbf{W^u}$ are also real analytic with respect to $\mathbf{X} \neq \mathbf{0}$. The version of the stable manifold theorem used here is referred to as the Hartman-Grobman Theorem \cite{Guckenheimer:2002}.  See Figure \ref{fig:2}.

\medskip
 
It is remarked that the stable manifold theorem is applicable to a more general set of differential equations for which (\ref{eq:NonlinearSystem}) is a special case,
\begin{equation}
\mathbf{X}\rq{} = \mathbf{CX} + \mathbf{g(X)},
\label{eq:System1}
\end{equation}
where $\rq{} \equiv d/d\tau$, $\tau \in \mathbb{R}$, $\mathbf{X} \in \mathbb{R}^n$, and $\mathbf{C}$ is an $n \times n$ real constant matrix. It is assumed all the eigenvalues of $\mathbf{C}$ are purely real and non-zero. We assume $\mathbf{g(0)} = \mathbf{0}$, and ${{\partial \mathbf{ g} }\over{\partial \mathbf{ X}} }(\mathbf{0})$ = $\mathbf{0} $ (${{\partial \mathbf{ g} }\over{\partial \mathbf{ X}} }$ is an $n \times n$ matrix) The linearized system is
\begin{equation}
\mathbf{\bar{X}}\rq{} = \mathbf{C\bar{X}}.
\label{eq:System2}
\end{equation}
If $\mathbf{g(X)}$ is $C^k$, $k \geq 1$, or real analytic, $\mathbf{g(0)} = \mathbf{0}$, and ${{\partial \mathbf{g} }\over{\partial \mathbf{ X}} }(\mathbf{0})$ = $\mathbf{0} $ (${{\partial \mathbf{g} }\over{\partial \mathbf{ X}} }$ is a matrix), then the stable manifold theorem states that there exists a homeomorphism $\mathbf{h}$ between the trajectory points of  (\ref{eq:System1}) and  (\ref{eq:System2}) in a neighborhood of $\mathbf{X}$=$\mathbf{0}$ preserving the time parameterization. $\mathbf{X}(\tau)$ is $C^k$,  $k \geq 1$, or real analytic, respectively, with respect to $\tau$, and $\mathbf{W^s}, \mathbf{W^u}$ are $C^k$,  $k \geq 1$, or real analytic, respectively, with respect to $\mathbf{X}$.

\medskip\medskip\medskip\medskip

\noindent
 {\it Procedure for Branch Regularization}
\medskip\medskip

The branch regularization of the solutions is obtained by patching a given contracting solution $\mathbf{X}(t)$ for $t < 0$ to a unique expanding solution ${X}(t)$ for $t >0$ at $t=0$ for $\mathbf{X}=\mathbf{0}$ in a continuous manner for the nonlinear system, after application of the stable manifold theorem.
 
The patching procedure proves that as $\tau \rightarrow \infty$ in the contracting universe, where $\mathbf{X}(\tau) \rightarrow 0$, and equivalently $t \rightarrow 0^-$, $a(t)$ satisfies
\begin{equation}
a(t) = (-t)^{2 /( 3 (w_c + 1))}\Psi(t),
\label{eq:AofT}
\end{equation}
where $\Psi(t)$ is a real analytic function defined for positive and negative $t$ in an open neighborhood of $0$, and where $\Psi(0) \neq 0$.

This equation implies that $a(t)$ is a well defined real number for $t > 0$ if and only if  $w_c \in \mathbb{P}_w$. The necessity and sufficiency follows from its derivation. It is noted that for $t < 0$ in the contracting universe we are using (\ref{eq:TimeMap}) which implies (\ref{eq:NTsolution}). When going to an expanding universe for $t > 0$ the minus sign is replaced by a plus sign in (\ref{eq:TimeMap}),  (\ref{eq:NTsolution}), implying $\tau \rightarrow -\infty$ as $ t \rightarrow 0^+$.
\medskip

The unique extension of $a(t)$, together with the regularized system of differential equations \ref{eq:dQdN}, \ref{eq:dwdN}, \ref{eq:dOmegadN}, proves that the variables $Q(t), W(t), \Omega(t)$, and also $H(t)$ as noted earlier,  have a unique branch extension from $t \leq 0$ to $t \geq 0$.

It is noted that showing $a(t)$ can be extended through $t=0$ implies that $\tau(t)$ can be continued through $t=0$. The extension of $\tau$ is well defined, as with $H$, even though they are not defined at $t=0$ and switch sign.

We briefly sketch the idea of the  proof of (\ref{eq:AofT}).  It was first proven in \cite{Belbruno:2013} where all energy components have constant equations of state. Its application to the current modeling  is described in \cite{Xue:2014}, summarized here. If one considers the basis of eigenvector solutions (\ref{eq:Eigenvectors}) for $\bar{Q}(\tau), \bar{W}(\tau), \bar{\Omega}(\tau)$  it can be explicitly determined that (\ref{eq:AofT}) without $\Psi(t)$ is a solution, $\bar{a}(t)$, of the differential equation
\begin{equation}
\ddot{\bar{a}} = - \frac{(3 w_c + 1)}{2 \bar{a}^{3 w_c + 2}}.
\label{eq:LinearA}
\end{equation}
For the dynamical system for $Q(\tau), W(\tau), \Omega(\tau)$, it can be shown that $a(t)$ satisfies 
\begin{equation}
\ddot{a} = - \frac{(3 w_c + 1)}{2 a^{3 w_c + 2}} - f(a) ,
\label{eq:NonlinearA}
\end{equation}
where $f(a)$ is subdominant with respect to the leading order term for $a$ small.

It is shown in \cite{Xue:2014} that the trajectory points of  (\ref{eq:LinearA}) can be homeomorphically mapped onto the trajectories of (\ref{eq:NonlinearA}) for $a$ sufficiently small. This follows by using an argument described in \cite{Belbruno:2013}, which, in turn, is based on a version of the stable manifold theorem used in \cite{McGehee:1981} (Lemma 7.5).  This implies that the trajectory points of the linear system (\ref{eq:LinearA}) can be continuously mapped 1-1 onto the trajectory points of  (\ref{eq:NonlinearA}). This proves that  $w_c \in \mathbb{P}$ is required for the nonlinear system.  This also proves that $t$ continuously moves from $t\leq 0$ to $t \geq 0$ yielding a unique branch extension.
\medskip

\noindent
\textit{Remark } The results in  Reference~\cite{Belbruno:2013} yield the same conditions as in Theorem ~1, with $w_c \in \mathbb{P}_w$, in the more restricted case with constant $w(t) = w_c$, or $W(t) = 0$, and for any additional energy components with constant equation of state.  That is, only $a(t)$ varies. 
\medskip

The results explained in this section are deterministic in nature in the sense that the differential equations have well defined terms and the solutions have well defined values for each value of time. In the next section the situation is different where random perturbations are included. In this case, different methods are needed.

\section{Stochastic Modeling and Definitions} \label{sec:3}

In this section we consider the case where random perturbations are added to the original System \ref{eq:DiffEqusMoreGeneralH}, \ref{eq:DiffEqusMoreGeneralW}, \ref{eq:DiffEqusMoreGeneralOmega} in the time variable $t$.  We will model this system as a It\^{o} stochastic system and the perturbations are modeled as Brownian motions.
\medskip

We begin by considering System \ref{eq:DiffEqusMoreGeneralH}, \ref{eq:DiffEqusMoreGeneralW}, \ref{eq:DiffEqusMoreGeneralOmega} which we write as
\begin{equation}
{{d\mathbf{Y}}\over{dt}}  = \mathbf{F}(\mathbf{Y}),
\label{eq:OriginalSystem}
\end{equation}
where $\mathbf{Y} = (Y_1, Y_2, Y_3) = (H, W, \Omega)$, and $\mathbf{F}(\mathbf{Y}) = (F_1(\mathbf{Y}), F_2(\mathbf{Y}), F_3(\mathbf{Y})) $ equals the right hand side of \ref{eq:DiffEqusMoreGeneralH}, \ref{eq:DiffEqusMoreGeneralW}, \ref{eq:DiffEqusMoreGeneralOmega} , respectively. This is the original system of differential equations in the time variable $t$ whose solution is denoted by $\mathbf{Y}(t)$. The big bang is at $t=0$. $\mathbf{Y}(0) =  (-\infty, 0, 0)$ from the contracting universe, where $t \rightarrow 0^-$ (in the case of the expanding universe, where $t \rightarrow 0^+$, $\mathbf{Y}(0) = (\infty, 0, 0)$).

We introduce random perturbations into this system by the addition of the term  $\mathbf{\tilde{R}}$ on the right hand side, 
\begin{equation}
{{d\mathbf{Y}}\over{dt}}  = \mathbf{F}(\mathbf{Y}) + \mathbf{\tilde{R}}(\mathbf{Y}, t),
\label{eq:OriginalSystem2}
\end{equation}
where
\begin{equation}
\mathbf{\tilde{R}}(\mathbf{Y}, t) = \sum_{i=1}^m \mathbf{G}_i(\mathbf{Y}){{d\tilde{\mathcal{W}}_i(t)}\over{dt}}.
\label{eq:R}
\end{equation}
The $\tilde{\mathcal{W}}_i(t), i= 1,..., m$ are one-dimensional Brownian motions, which are defined as stochastic processes, equivalently called Wiener processes. $m \geq 1$ is arbitrary. Their derivative $d\tilde{\mathcal{W}}_i(t) /dt $ is formally defined as white noise functions $\xi_i(t)$ which have a constant spectral density.  The $\mathbf{G}_i(\mathbf{Y})$ are arbitrary smooth functions whose regularity and properties are defined below.

To understand the behavior of System \ref{eq:OriginalSystem2} due to the random terms $\xi_i$, $i=1,..., m$,  it is transformed to a special stochastic system of differential equations, called an  {\it It\^{o} system}.  To do this,  we view $\mathbf{Y}$ as a three-dimensional random variable defined on a probability space $\mathcal{P}$.  The probability space $\mathcal{P}$  is denoted by the triple $(\tilde{\Omega}, \mathcal{U}, P)$ consisting of a nonempty set $\tilde{\Omega}$, the sample space, an arbitrary subset $\mathcal{U}$ of $\tilde{\Omega}$ (the subsets $\mathcal{U}$ form a sigma algebra), and a probability measure $P$, $P: \mathcal{U} \rightarrow [0,1]$. The probability that $ \mathbf{Y} \in \mathcal{U}$, due to the perturbations of the Brownian motion within the term $\mathbf{\tilde{R}}$, is $P(\mathcal{ U})$. 

Of particular importance to this study is the change of $\mathbf{Y(\omega)}$ as a function of time $t \geq 0$, $\omega \in \tilde{\Omega}$.  The set $\{ \mathbf{Y}(t)| |t| \geq 0\}$ is a stochastic process.   For each value of time, $t=t_1$, we can consider the distribution of values of $\mathbf{Y}(\omega)$ as $\omega$ varies. 

The It\^{o} system for (\ref{eq:OriginalSystem2}) is given by
\begin{equation}
{{d\mathbf{Y}}} = \mathbf{F}(\mathbf{Y})dt + \mathbf{{R}}(\mathbf{Y}, t),
\label{eq:OriginalSystemIto}
\end{equation}
where
\begin{equation}
\mathbf{{R}}(\mathbf{Y}, t) = \sum_{i=1}^m \mathbf{G}_i(\mathbf{Y}){{d\tilde{\mathcal{W}}_i(t)}}.
\label{eq:RIto}
\end{equation}

This system for $t < 0$ is transformed to regularized coordinates $\mathbf{X}, \tau$ given by  (\ref{eq:HMap}), (\ref{eq:TimeMap}). It is verified that this yields
\begin{equation}
   d\mathbf{X} = \mathbf{B}(\mathbf{X})d\tau +  {\mathbf{\hat{R}}}({\mathbf{X}}) ,
\label{eq:ItoDE1}
\end{equation}
where, 
\begin{equation}
\mathbf{\hat{R}}(\mathbf{X}, \tau)  =  \sum_{i=1}^m(-Q)^{3/2}\mathbf{g}_i(\mathbf{X}) d\mathcal{W}_i(\tau), 
\label{eq:RT}
\end{equation}
\begin{equation}
 \mathbf{B}(\mathbf{X}) = \mathbf{A} \mathbf{X} + \mathbf{\Delta}(\mathbf{X}),
\label{eq:h(x)}
\end{equation}
$\mathbf{X} = (X_1,X_2, X_3) = (Q, W, \Omega) \in \mathbf{R^3}$ is a three-dimensional random variable, $\tau > 0$, $Q <0$ in the contracting universe. (In an expanding universe, $-Q$ is replaced by $Q$ in (\ref{eq:RT}).)  The matrix $\mathbf{A}$ and the vector $\Delta$ are given by (\ref{eq:MatrixA}), (\ref{eq:FCase3}).  $\mathbf{g}_i(\mathbf{X})$ is the vector function
$\mathbf{G}_i(\mathbf{Y})$ in regularized coordinates, $i=1,...,m$. $\mathbf{g}_i(\mathbf{X})$ are arbitrary  real valued vector functions of $\mathbf{X}$.  $\mathcal{W}_i$ are the Brownian motions in regularized time, $\tau$.  The sample path for $\mathbf{X}(\tau), \tau > 0$,    represents the solution to (\ref{eq:ItoDE1}).

The factor $(-Q)^{-3/2}$ occurs in $\mathbf{\hat{R}}$ due to the time transformation which gives a multiplicative factor -Q in $\mathbf{\hat{R}}$, together with the fact that for Brownian motion, $$d{\mathcal{W}}_i^2 = d\tau,$$ and the time transformation then implies $d{\tilde{\mathcal{W}}}_i(t) = (-Q)^{1/2}(d\tau)^{1/2} =  (-Q)^{1/2} d\mathcal{W}_i(\tau)$.
\medskip

Since $\mathbf{\hat{R}}(\mathbf{0}) = \mathbf{0}$ and $\mathbf{B(\mathbf{0})} = \mathbf{0}$, $\mathbf{X}(\tau) = \mathbf{0}$ is a stationary solution. This implies $\mathbf{X}(\tau) \rightarrow \mathbf{0}$ as $\tau \rightarrow \infty$ in the contracting universe, and $\mathbf{X}(\tau) \rightarrow \mathbf{0}$ as $\tau \rightarrow -\infty$ (backwards in time) for the expanding universe.
\medskip

It is noted that, since $\mathbf{B(\mathbf{0})} = \mathbf{0}$ and $  {\partial \Delta\over{\partial \mathbf{X}} }(\mathbf{0}) = \mathbf{0}$, we have ${\partial \mathbf{B}\over{\partial \mathbf{X}}}(\mathbf{0}) = \mathbf{A}$,   ${{\partial{\mathbf{\hat{R}}}}\over{\partial{\mathbf{X}}}}(\mathbf{0}) = \mathbf{\hat{0}}$ ($3 \times 3$ matrix of $0$).
\medskip

We summarize some of the properties  of  the It\^{o} system of differential equations (\ref{eq:ItoDE1}): 
\medskip \medskip

\noindent
Summary A - {\it Properties of the  It\^{o} System (\ref{eq:ItoDE1}) }
\medskip

\noindent
  i.) $\mathbf{B}(\mathbf{0}) = \mathbf{0}, {\partial \mathbf{B}\over{\partial \mathbf{X}}  }(\mathbf{0}) = \mathbf{A}$.     \\
\noindent
 ii.)  $\mathbf{B}(\mathbf{X})$ is real analytic function  for $\mathbf{X}$ in an open neighborhood of $\mathbf{0}$. \\ 
 \noindent
iii.) $\mathbf{X} = \mathbf{0}$ is a stationary solution. \\
 \noindent
 iv.) $\mathbf{\hat{R}}(\mathbf{0}) = \mathbf{0}$, ${{\partial{\mathbf{\hat{R}}}}\over{\partial{\mathbf{X}}}}(\mathbf{0}) = \mathbf{\hat{0}}$ ($3 \times 3$ matrix of $0$). \\
\noindent
  v.) $\mathbf{\hat{R}}(\mathbf{X}, \tau)$ is a $C^1$ function of $\mathbf{X}$  in an open neighborhood of $\mathbf{X} = 0$.  \\
\noindent
\medskip\medskip

It is assumed that at the big bang  $|\mathbf{\hat{g}}(0)| \neq 0$,  where $\mathbf{\hat{g}} = (\mathbf{g}_1, .... \mathbf{g}_m)$, a real valued $3 \times m$ matrix. This implies that at the big bang, $|\mathbf{G}(\mathbf{Y}(0))| \neq  0$. Thus, in the physical stochastic coordinates for the It\^{o} System (\ref{eq:OriginalSystemIto}), 
$|\mathbf{R}(\mathbf{Y}(0), 0)| \neq 0$.
\medskip

\medskip\medskip

\noindent
{\it Non-Zero Random Perturbations at the Big Bang}
\medskip

\noindent
In the physical coordinates $(\mathbf{Y}, t)$, $|\mathbf{R}| \neq 0$ at the big bang.  This implies that the random perturbations defined by the Brownian motion $\tilde{\mathcal{W}}_i, i=1,...,m,$ exist up to and including the big bang for the  It\^{o} System (\ref{eq:OriginalSystemIto}). 
\medskip\medskip\medskip\medskip

It is noted that  Equation \ref{eq:ItoDE1} in component form is given by
\begin{equation}
dX_j  = B_j(\mathbf{X})d\tau +  \sum_{i=1}^m (-Q)^{3/2}g_{ji}(\mathbf{X}) d\mathcal{W}_i, \hspace{.2in} j=1,2,3. 
\label{eq:ItoDE2}
\end{equation}
\medskip

The probability distribution for a general Brownian motion $\mathcal{W}(\omega)$, for a given value of $\tau$, is  normal  of  expectation $0$ and variance $\tau$, denoted by $N(0,\tau)$.   The probability, $P$, that $\mathcal{W}$ will take on values between $\alpha, \beta$ for $\alpha \leq \beta$, for all $\tau > 0$, is therefore given by
\begin{equation}
P(\alpha \leq \mathcal{W}(\tau) \leq \beta) = {1\over{\sqrt{2 \pi \tau}} } \int_{\alpha}^{\beta} e^{{-x^2}\over{2\tau}} dx.
\label{eq:WienerProb}
\end{equation}

It is seen that as $\tau \rightarrow \infty$, the variance, $V(\tau) \rightarrow \infty$.  This implies that for each $\tau$, the probability density function, $f_{\tau}(x) = (1/\sqrt{2\pi\tau}) e^{{-x^2}\over{2\tau}}$ flattens out and approaches $0$.
\medskip

$\mathcal{W}(\tau,\omega)$ has several interesting properties of the sample path worth noting. Let $\tau \rightarrow \mathcal{W}(\tau)$ be the sample path for  $\mathcal{W}(\tau,\omega)$ for $\tau > 0$. Then, $\mathcal{W}(\tau)$  is non-differentiable, has infinite variation(i.e. infinite length) on sub-intervals of $\tau$,  and is Holder continuous \footnote{i.e. for each $0 < \gamma < {{\alpha}\over{\beta}}$, $\hat\tau > 0$ and a.e. $\omega$,  there exists a constant $K=K(\omega, \gamma,\hat\tau)$ such that $|\mathcal{W}(\tau,\omega) - \mathcal{W}(s, \omega)| \leq K|\tau-s|^{\gamma} $ for all 
$0 \leq s,\tau \leq \hat\tau $ where $ \beta =2 m, \alpha = m-1,  m=2,3,\ldots  \hspace{.2in} {{\alpha}\over{\beta}} = {1\over2} - {1\over{2m}} $. } in $\tau$ a.e.in $\omega$.
\medskip


\section{Stochastic Regularization} \label{sec:4}

We show that stochastic branch regularization can be done for (\ref{eq:OriginalSystemIto}) at $\mathbf{Y} = \mathbf{Y}(0)$ for the same co-prime conditions as in Theorem 1. This is accomplished by using a stochastic version of the stable manifold theorem \cite{Mohammed:1999}. The results are summarized in Theorem 2.
\medskip\medskip

To do this we consider the regularized  It\^{o} System of stochastic differential equations given by (\ref{eq:ItoDE1}). We will be considering $\bf{X}$ near to $\mathbf{0}$ and considering first a contracting universe.
\medskip\medskip

The local existence and uniqueness of solutions $\mathbf{X}(\tau)$ with an initial point $\mathbf{X_0}$ for $\tau = \tau_0 > 0$ are guaranteed by only requiring that  $\mathbf{B(\mathbf{X})}$,  $\mathbf{\tilde{g}}_i(\mathbf{X}) \equiv (-Q)^{3/2}\mathbf{g}_i(\mathbf{X})$, $i=1,...,m$ are Lipschitz continuous in a small neighborhood of $\tau_0$, yielding  continuous, but non-differentiable sample paths. This is satisfied for  $\mathbf{B(\mathbf{X})}$ and  $\mathbf{\tilde{g}}_i(\mathbf{X})$ since they are real analytic for $\tau_0 < \tau<  \infty$.

It is important to note that a solution $\mathbf{X}(\tau)$ of (\ref{eq:ItoDE1}) depends on $\mathbf{X_0}$ as well as $\tilde{\Omega}$ due to the random process $\mathcal{W}(\tau)$. This implies that more generally, The solution is written as $\mathbf{X}(\tau, \mathbf{X_0}, \omega)$, where $\omega \in \tilde{\Omega}$.  Thus,  $\mathbf{X}(\tau, \mathbf{X_0}, \omega) : \mathbf{R^1} \times \mathbf{R^3} \times \tilde{\Omega} \rightarrow \mathbf{R^3}$.  For simplicity of notation we write  $\mathbf{X}(\tau,  \omega)$ or just $\mathbf{X}(\tau)$.

It is remarked that since $\bf{X}$ is a random variable, when we say that $\mathbf{X}(\tau)$ is locally unique, taking on the initial value $\mathbf{X_0}$, this uniqueness is in probability almost surely (a.s.).


The theorem we will prove is:
\medskip\medskip

\noindent
\textbf{Theorem~2.} \; The solutions to the classical Friedmann equations for $H, W, \Omega$ with random, small perturbations up to and including the big bang, modeled as an It\^{o} system with Brownian motion given by  (\ref{eq:OriginalSystemIto}), can be uniquely extended through the big bang at $t=0$ as a branch regularization  if and only if the value of $w_c$ belongs to the discrete set $\mathbb{P}_w$ of co-prime values of the equation of state.  The curves of the branch solutions are continuous but non-differentiable.

\medskip\medskip\medskip
\noindent
{\it Physical Description of Motion}
\medskip

\begin{figure}[h!]
\centering
	\includegraphics[width=0.65\textwidth, clip, keepaspectratio]{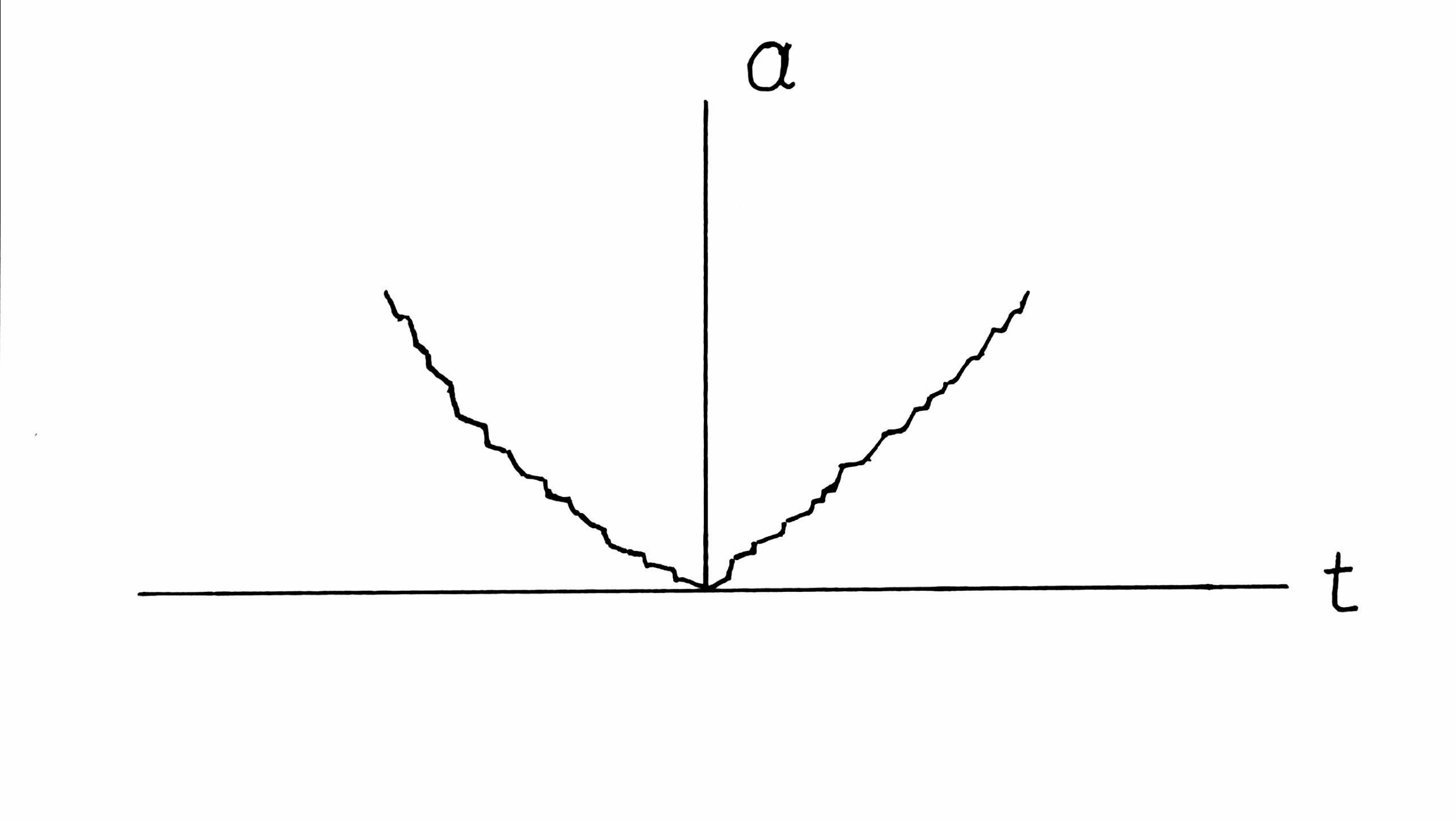}
\caption{Stochastic behavior of $a(t)$ in a neighborhood of $t=0$ where $w = w_c \in \mathbb{P}_w.$ }
\label{fig:3}
\end{figure}

\noindent
The solutions of the randomly perturbed Friedmann equations using the stochastic model will be different from the classical ones found in \cite{Xue:2014}, described in Section \ref{sec:2}. For example the curve for  $(t, a(t))$ for $t$ near $0$ will no longer be smooth. In this case, it will be non-differentiable. This means that locally it will move in a Brownian fashion.  (See Figure \ref{fig:3}.)
This is also true for $Q(t), w(t), \Omega(t)$.  It is recalled that $\mathcal{W}(\tau)$ also has infinite variation, which adds considerable complexity to the solutions. A striking aspect of this result is that the original system in physical coordinates has random perturbations also at the big bang. However, although there are random perturbations at the big bang, $a(t) \rightarrow 0$ and $w \rightarrow w_c$ as $|t| \rightarrow 0$ due to the form of (\ref{eq:AofTStochastic}).
\medskip\medskip\medskip

\noindent
Proof of Theorem 2.
\medskip\medskip\medskip\medskip

The proof of this theorem parallels the proof of Theorem 1 described in Section \ref{sec:2}.
\medskip

\noindent {\it Step 1.}  \hspace{.1in}
$\bf{X} = \mathbf{0}$ is a stationary point of  (\ref{eq:ItoDE1}).  The conditions ${{\partial{\mathbf{\hat{R}}}}\over{\partial{\mathbf{X}}}}(\mathbf{0}) = \mathbf{\hat{0}}$  and ${\partial \mathbf{B}\over{\partial \mathbf{X}}  }(\mathbf{0}) = \mathbf{A}$, listed in Summary A,  imply that the linearized system about $\mathbf{X}=\mathbf{0}$ is given by
\begin{equation}
  d\mathbf{\bar{X}}(\tau) = \mathbf{A}\mathbf{\bar{X}}(\tau)d\tau.
\label{eq:ItoLinear}
\end{equation}
Since the stochastic term is not included, this can be solved in the classical manner as was done in Section \ref{sec:2}. This yields the eigensolutions (\ref{eq:Eigenvectors}). The Lyapunov exponents , or eigenvalues,  $\lambda_i < 0, i=1,2,3$ of  $\mathbf{A}$ are all negative and not equal in the contracting universe where $\tau < 0$. Thus, depending on the value of $w_c$, they can be ordered as
$ \lambda_{i_1}< \lambda_{i_2} < \lambda_{i_3} < 0$, where $i_j \neq i_k, j \neq k$, $i_1, i_2, i_3$ is some ordering of $1,2,3$. (Likewise, in an expanding universe where $\tau < 0$, as described in Section \ref{sec:2}, the eigenvalues are all positive, which we order as  $0 < -\lambda_{i_3}< -\lambda_{i_2} < -\lambda_{i_1}$). This implies that $\mathbf{X} = \mathbf{0}$ is a hyperbolic stationary point of (\ref{eq:ItoDE1}). This follows using the definition in \cite{Mohammed:1999} since the eigenvalues are real, non-zero and distinct.
The definition in \cite{Mohammed:1999} is for the more general case of a hyperbolic stationary trajectory defined as a stochastic process, which includes the special case of a point.
\medskip

\noindent{\it Step 2.}   \hspace{.1in} Apply a stochastic version of the stable manifold theorem. The general stochastic version of the stable manifold theorem is stated as Theorem 3.1 in \cite{Mohammed:1999}.  It is formulated for a far more general modeling than we have here. However, Corollary 3.1.1 of \cite{Mohammed:1999} is directly applicable to our model. We state this result for our modeling.
\medskip\medskip

\noindent Stochastic Stable Manifold Theorem (Mohammad-Scheutzow \cite{Mohammed:1999}). \hspace{.1in}  Consider the It\^{o} system of stochastic differential equations \ref{eq:ItoDE1}, where $\mathbf{B}(\mathbf{0}) = \mathbf{0}$,  $\mathbf{\hat{R}}(\mathbf{0}) = \mathbf{0}$,
${\partial \mathbf{B}\over{\partial \mathbf{X}}  }(\mathbf{0}) = \mathbf{A}$, ${ {\partial{\mathbf{\hat{R}}} } \over { \partial{\mathbf{X}} } } (\mathbf{0}) = \mathbf{\hat{0}}$ and where the Lyapunov exponents of the linearized flow (\ref{eq:ItoLinear}) are given by $\lambda_{i_k}, k=1,2,3$, and $\mathbf{X} = \mathbf{0}$ is a hyperbolic stationary point. Also, where $\mathbf{B}(\mathbf{X}) , \mathbf{{g}}_i(\mathbf{X}) $ are $C^{1}$ functions of $\mathbf{X}$ for $\mathbf{X}$ in an open neighborhood $\mathbf{0}$. Let $\mathbf{\bar{W}^{s,u}}$ be the stable and unstable invariant manifolds to $\mathbf{X} = \mathbf{0}$ for (\ref{eq:ItoLinear}) as obtained in Section \ref{sec:2}. Then there exists
\medskip

\noindent
i.) Stable and unstable manifolds to $\mathbf{X} = \mathbf{0}$ for (\ref{eq:ItoDE1}), lableled $\mathbf{W^{s,u}}(\omega)$, which are $C^{1}$, \\
\noindent
ii.)  Continuous and non-differentiable solutions  $\mathbf{X}(\tau,  \omega)$, \\
\noindent
iii.) $\mathbf{X}(\tau,  \omega) \rightarrow \mathbf{0}$ as $\tau \rightarrow \infty$ for $\mathbf{X_0}\in \mathbf{W^{s}}(\omega)$ and  $\mathbf{X}(\tau,  \omega) \rightarrow \mathbf{0}$ as $\tau \rightarrow -\infty$ for $\mathbf{X_0}\in \mathbf{W^{u}}(\omega)$, \\
\noindent
iv.) $\mathbf{W^{s,u}}(\omega)$ are invariant for the flow  $\mathbf{X}(\tau,  \omega) $ for $|\tau|$ sufficiently large, \\
\noindent
v.)  For $\mathbf{X_0} \in \mathbf{W^{s}}(\omega)$ 
 \begin{equation}
|\mathbf{X}(n, \mathbf{X_0}, \omega)|  \leq \beta_1(\omega) e^{(\lambda_{i_0} + \epsilon_1) n },
\label{eq:ExpEst}
\end{equation}
for all integers $n \geq 0$, where $ \lambda_{i_0} = \max \{ \lambda_{i_k} \}, k=1,2,3$, $\epsilon_1 \in (0, -\lambda_{i_0})$ is fixed, $\beta_1(\omega)$ is a random variable, and
\noindent 
\begin{equation}
\lim \sup_{\tau \rightarrow \infty}   \log |\mathbf{X}(\tau, \mathbf{X_0}, \omega)| \leq \lambda_{i_0}\\
\label{eq:Log}
\end{equation}
\noindent
vi.)  For $\mathbf{X_0} \in \mathbf{W^{u}}(\omega)$ 
 \begin{equation}
|\mathbf{X}(-n, \mathbf{X_0}, \omega)|  \leq \beta_2(\omega) e^{(-\lambda_{i_0-1} + \epsilon_2) n },
\label{eq:ExpEst2}
\end{equation}
for all integers $n \geq 0$, where $ \lambda_{i_0-1} = \min \{ -\lambda_{i_k} \}, k=1,2,3$, $\epsilon_2 \in (0, \lambda_{i_0 - 1})$ is fixed, $\beta_2(\omega)$ is a positive random variable
\noindent 
\begin{equation}
\lim \sup_{\tau \rightarrow \infty} {1\over{\tau}}  \log |\mathbf{X}(\tau, \mathbf{X_0}, \omega)| \leq -\lambda_{i_0 - 1}\\
\label{eq:Log2}
\end{equation}
\noindent
vii.)  $\mathbf{W^{s}}(\omega)$ is tangent to  $\mathbf{\bar{W}^{s}}$ at $\mathbf{X} = \mathbf{0}$ and $\mathbf{W^{u}}(\omega)$ is tangent to  $\mathbf{\bar{W}^{u}}$ at $\mathbf{X} = \mathbf{0}$, \\
\noindent
viii.) $\mathbf{W^{s}}(\omega)$ is transversal to $\mathbf{W^{u}}(\omega)$ at $\mathbf{X} = \mathbf{0}$. \\
\noindent

It is interesting to note that unlike the classical stable manifold theorem where one has exponential decrease (or increase) of the solution to (away from) the stationary point for any time $\tau$, in this case, as is seen by (\ref{eq:ExpEst}), (\ref{eq:ExpEst2}), this is true only for integer values $n$. This is due to the random behavior of $\mathcal{W}_i$. However, this is not true for the log estimates, (\ref{eq:Log}), (\ref{eq:Log2}).

From this theorem, we can conclude, in particular, that in a contracting universe, as $\tau \rightarrow \infty$, $\mathbf{X}(\tau, \mathbf{X_0}, \omega) \rightarrow  \mathbf{0}$ in an approximately exponential manner corresponding to integer time values. Likewise, in an expanding universe, $\tau \rightarrow -\infty$, $\mathbf{X}(\tau, \mathbf{X_0}, \omega) \rightarrow  \mathbf{0}$ also in an approximately exponential manner for integer time values.

Thus, if we consider a particular solution, $\mathbf{X}(\tau)$, approaching $\mathbf{X}=\mathbf{0}$ in a contracting universe, we can piece it together with a solution at $\mathbf{X} = \mathbf{0}$ for an expanding universe in the same manner as we did in Section \ref{sec:2}. The uniqueness of this continuous extension at $t=0$ follows as described at the end of Section \ref{sec:2} where the same argument can be used to also get the continuous extension in time through $t=0$ only for $w_c \in \mathbb{P}_w$.   There are some modifications that must be done to that argument. When applying the stochastic version of the stable manifold  theorem, the differential equation (\ref{eq:NonlinearA}) for $a(t)$, must first  be put in a stochastic framework by making an It\^{o} system analogous to what was done here.  When this is done, it implies that $a(t)$ becomes a stochastic process due to the addition of a Brownian motion, satisfying (\ref{eq:NonlinearA}) with a random perturbation term added to the left hand side. This is done as follows:
\medskip

\noindent
Since the flow of  (\ref{eq:NonlinearA}) is equivalent to the flow of the unperturbed equation (\ref{eq:LinearA}) in a neighborhood of the the big bang at $a=0$ and $t=0$ by a special version of the  stable manifold theorem that was described previously, it suffices to consider (\ref{eq:LinearA}).  As is shown in \cite{Belbruno:2013}(see Theorem 5, Equation 26), by transforming  (\ref{eq:LinearA})  with a McGehee transformation of $a, da/dt, t$ to $r, v, s$, respectively ($a = r^{\gamma}, da/dt = r^{-\beta\gamma}v, dt = r ds$), and translating  $v \rightarrow v - c$, then  the big bang occurs at $r = 0 , v = 0$ and $s=0$.  The differential equations for $r,v$ can be written as, by expanding in a Taylor series about $r=0, v=0$,
\begin{equation}
{d{\mathbf{x}}\over{ds}} = \mathbf{K} \mathbf{x} + \mathbf{d}(\mathbf{x}),
\label{eq:LinearDEs}
\end{equation}
where, $\mathbf{x} = (r,v)$, 
\begin{equation}
\mathbf{K} = \left( \begin{array}{cc}
(\beta +1)c& 0  \\[4pt]
0 & 2\beta c   \end{array} \right) ,
\label{eq:MatrixK}
\end{equation}
\begin{equation}
\mathbf{d}(\mathbf{x}) =  \left( \begin{array}{c}
(\beta +1)rv,  \\[4pt]
\beta v^2   \end{array} \right)
\end{equation}
$c = \pm\sqrt{2}$ (+ is for an expanding universe, - is for a contracting universe), $\beta = \alpha/2, \alpha = 3(1+w_c)-2, \gamma = (1+\beta)^{-1}$, $\beta > 2$ (corresponding to $w_c > 1), 0 < \gamma < 1/3.$

It is seen that the big bang $r=0, v=0$ is a hyperbolic rest point for (\ref{eq:LinearDEs}), where $ \mathbf{x}(s) \rightarrow \mathbf{0}$ as $|s| \rightarrow \infty$.  Since ${\partial  {\mathbf{d}}\over{\partial \mathbf{\mathbf{x}}} }(\mathbf{0}) = \mathbf{\tilde{0}}$, where $ \mathbf{\tilde{0}} $ is a zero $2 \times 2$ matrix, then the linearized solutions of (\ref{eq:LinearDEs}) are spanned by the basis solutions, $(e^{(\beta+1) c s}, 0), (0, e^{2\beta c s )})$, and we have $t = \int (e^{(\beta+1) c s}) ds = ((\beta+1) c)^{-1}e^{(\beta+1) c s}$, which implies $s = ((\beta+1)c)^{-1} \ln |t|$, implying $r \propto t$. Thus $a = r^\gamma \propto t^\gamma$ implies (\ref{eq:AofT}) is satisfied in the linearized case, where $t$ is replaced by $-t$ for the contracting universe.

An alternate way of obtaining (\ref{eq:AofT}) in the perturbed case is to use the more general stable manifold theorem applied to (\ref{eq:NonlinearA}) instead of the special version in \cite{McGehee:1981}(Lemma 7.5), described previously. In this case, the right hand side (\ref{eq:LinearDEs}) has an additional term $\mathbf{P}(\mathbf{x})$ representing the  perturbations due to the function $f(a)$ in (\ref{eq:NonlinearA}) after application of the McGehee transformation. This implies that $r \propto t G(t)$ for some function $G(t)$. This then implies that  $a = r^\gamma = t^\gamma \tilde{G}$, where $\tilde{G} = G^\gamma$.  This is of the form of  (\ref{eq:AofT}).

Viewing $\bf{x}$ as a two-dimensional random variable and $\bf{x}$ as a stochastic process, we  form an  It\^{o}  system from (\ref{eq:LinearDEs}),
\begin{equation}
d\mathbf{x} = \mathbf{K}(\mathbf{x})ds +  {\mathbf{{\Phi}}}({\mathbf{x}},s) ,
\label{eq:ItoDE3}
\end{equation}
where,
\begin{equation}
\mathbf{\Phi}(\mathbf{x}, s)  =  \sum_{i=1}^m\mathbf{\hat{g}}_i(\mathbf{x}) d\mathcal{W}_i(s).
\label{eq:R2}
\end{equation}
It is assumed that $\hat{g}_i(\mathbf{x})$ are $C^1$ functions of $\mathbf{x}$ in a neighborhood of $\mathbf{x} = \mathbf{0}$, $\mathbf{\hat{g}}_i(0) = \mathbf{0}$, and ${ {\partial \mathbf{\hat{g}}_i }\over{\partial \mathbf{x}} }(\mathbf(0)) =0$.  It is verified that this stochastic system satisfies all the necessary conditions for the stochastic stable manifold theorem.  In particular, this implies after application of the inverse McGehee regularization, $s \rightarrow t, r \rightarrow a$, we obtain
\begin{equation}
a(t) = (-t)^{2 / (3 (w_c + 1))}\tilde \Psi(t) 
\label{eq:AofTStochastic}
\end{equation}
in place of (\ref{eq:AofT}), where $\tilde{\Psi}(t)$ is a non-differentiable function, $\tilde{\Psi}(0) \neq 0$, and $\tilde{\Psi}(t)$ is defined for both $t \leq 0$ and $ t \geq 0$ (for $t \geq 0$, $-t$ is replace by $t$ in (\ref{eq:AofTStochastic}) ).

This yields the proof of Theorem 2.
\medskip\medskip\medskip\medskip

\noindent
{\it Properties of $a(t)$}
\medskip

\noindent
Although $\tilde{\Psi}(t)$ is non-differentiable,  it is also monotonic decreasing for $t \leq 0$ (increasing for $t \geq 0$.) and $|t|$ near $0$. This yields a monotonic, random, non-smooth, jagged variation for $a(t)$.  We consider $t \leq 0$.  This is deduced from the fact that $ a = e^{-\tau}$, where $\tau$  is large. This implies $a$ as a function of $\tau$ is monotone decreasing.  We also have $dt = (-Q) d\tau$, where $-Q > 0$ and therefore, $dt/d\tau > 0$. This implies $da/dt = (da/d\tau) (d\tau/dt) < 0 $, and hence $a(t)$ is monotone decreasing as $t \rightarrow 0$.  The variation of $a(t)$ is seen to decrease in a random, non-uniform and non-differentiable manner as $t \rightarrow 0$ yielding a non-smooth, jagged variation.  The same behavior of $a(t)$ is obtained for an expanding universe, where, in this case, $a(t)$ increases, and where $Q$ is replaced by $-Q$, $\tau$  by $-\tau$, $t$ by $-t$. See Figure \ref{fig:3}.

\medskip\medskip\medskip

\section{Numerical Verification} \label{sec:n}

The result that $\mathbf{X}(\tau, \omega) \to \mathbf{0}$ as $\tau \to \infty$, stated by the Stochastic Stable Manifold Theorem, is tested by numerical simulation of the stochastic dynamical system (\ref{eq:ItoDE1}). For simplicity, we take the matrix $g_{ji}(\mathbf{X})$ to be constant and diagonal, $g_{ji}(\mathbf{X}) = \delta_{ji}$, the latter being the Kronecker delta. Then the equation becomes, in component form,
\begin{equation} \label{eq:sim}
dX_i = B_i(\mathbf{X}) d\tau + (-Q)^{3/2} d\mathcal{W}_i , \quad i = 1, 2, 3.
\end{equation}
This system of It\^{o} stochastic differential equations is integrated using the Euler-Maruyama method \cite{Kloeden:1992}.

For a contracting universe with $t < 0$, the values of the variables should satisfy $Q < 0$, $W > 1 - w_c$, and $0 < \Omega < 1$. Furthermore, for the Stochastic Stable Manifold Theorem to hold, they must be in a small neighborhood of $\mathbf{0}$.

\begin{figure}
\centering
\includegraphics[width=\textwidth]{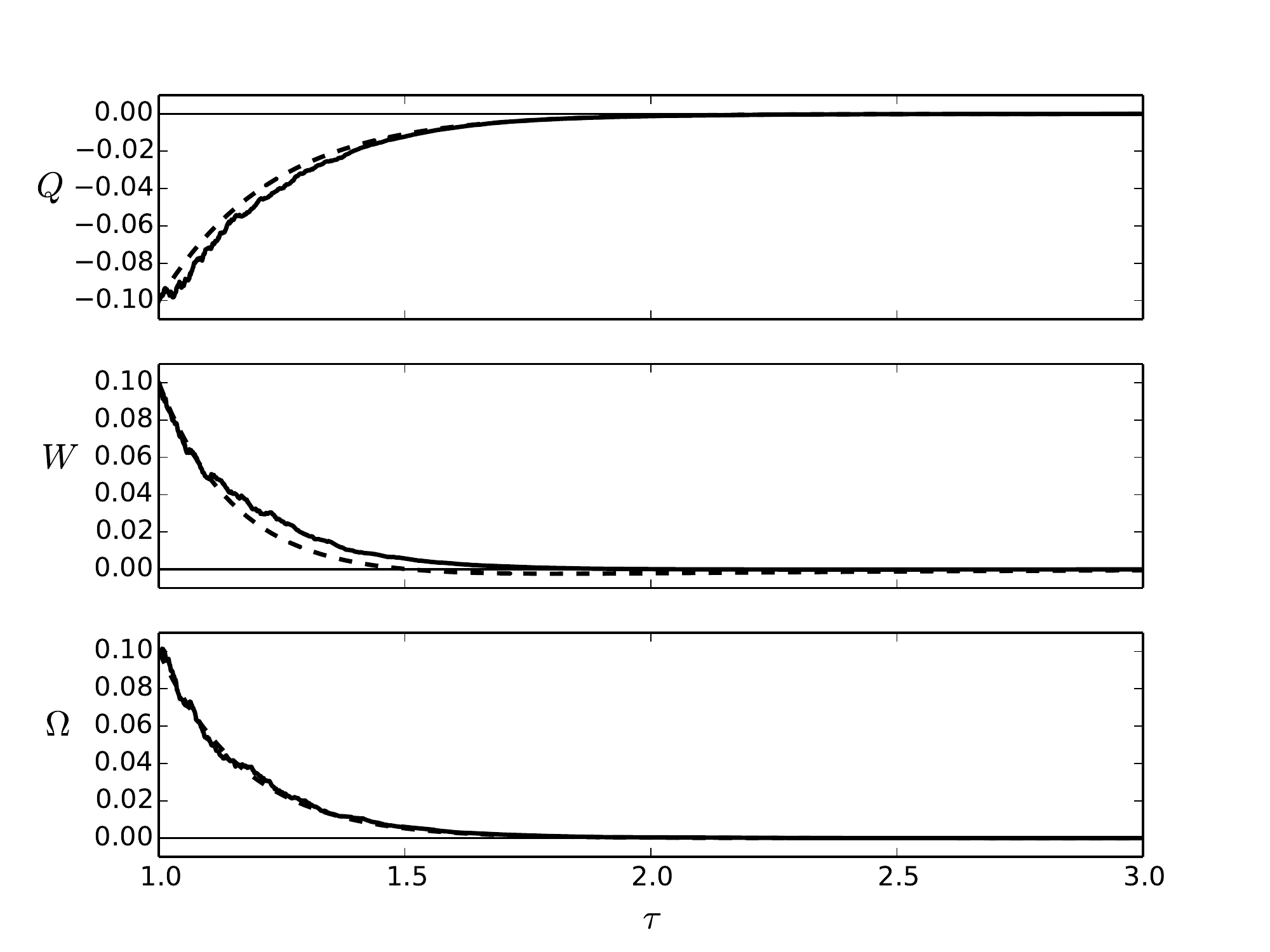}
\caption{Numerical simulation of the stochastic differential equations (\ref{eq:sim}), showing solutions with (solid) and without (dashed) random perturbations.}
\label{fig:bang}
\end{figure}

As an example, we choose $(Q_0, W_0, \Omega_0) = (-0.1, 0.1, 0.1)$, and start from $\tau_0 = 1$. The equations of state are chosen to be $w_c = 2$ and $w_1 = 0$. Fig.~\ref{fig:bang} shows a typical run of the simulation. It can be seen that the stochastic solution converges to $\mathbf{0}$ as well as the solution without random perturbations.

This result demonstrates the fact that if the solution starts sufficiently close to the big bang, the stochastic solutions remain close to the original solutions and converge to $\mathbf{0}$ , which is the main idea of the proof of Theorem 2. It is important to start the stochastic solution sufficiently near the big bang, otherwise, it may not stay close to the original solution by violating required bounds for the variables due to the stochastic fluctuations. 

It is noted that the transformed time variable $\tau$ goes to $\infty$ as the real time $t \to 0^-$. The solution for an expanding universe with $t > 0$ is similar but time-reversed, replacing $\tau$ by $-\tau$. Theorem 2 states that a contracting solution can be pieced together with an expanding solution at the big bang at $t = 0$.

\section{Discussion of Results} \label{sec:5}

The curve of $a(t)$ near the big bang has a random, non-smooth, jagged or choppy  appearance. This is summarized in:
\medskip

\noindent
\textit{Random, Monotonic, Non-smooth, Variations of Scale Factor}
\medskip

\noindent
The random, monotonic, non-differentiable, non-smooth variations of $a(t)$ for $a$ near $0$ are shown in Figure \ref{fig:3}.  The behavior of $a(t)$ occurs sufficiently near the big bang as required in the proof of Theorem 2. To determine how close is out of the scope of this paper. It is also noted that although there are random perturbations at the big bang, Equation \ref{eq:AofTStochastic} implies that $a(t) \rightarrow 0$ as $|t| \rightarrow 0$. 
\medskip\medskip

The non-differentiability and random nature of the step like behavior may give rise to gravity waves. This would require an analysis that is out of the scope of this paper, and is a topic for further study. 
\medskip\medskip

The addition of stochastic perturbations to the Friedmann differential equations yield complex perturbations, where it may be possible to model them as quantum fluctuations:
\medskip\medskip

\noindent
{\it Quantum Fluctuations}
\medskip

\noindent
By suitable choice of $\mathbf{G}_i(\mathbf{Y})$, $i=1,...,m$ for $m$ sufficiently large, it may be possible to model quantum fluctuations by the $\mathbf{R}(\mathbf{Y},t)$ at each time $t$. This is not analyzed here, and is a topic for further study. 
\medskip\medskip

The stochastic perturbations provide a general approach to perturb the Friedmann  equations. The term given by  $\mathbf{R}(\mathbf{Y},t)$ is general in nature, where one has the freedom to choose the functions as desired to model infinitely many different types of perturbations, where they need not be known in general. 
\medskip\medskip

The perturbed system of differential equations modifies the vector field of the unperturbed system. This does not alter the homogeneous metric we are using, given by (\ref{eq:metric}). However, it may be possible to interpret the perturbed equations as being due to a perturbed metric. This situation is summarized in:
\medskip\medskip

\noindent
{\it Spatial Inhomogeneity}
\medskip

\noindent
The model studied in this paper assumes spatial homogeneity, as true for the Friedmann equations. However, the random perturbations introduced in the model could potentially also represent local perturbations in space. One could consider different realizations of the random perturbations as happening in different parts of the universe.

In \cite{Xue:2013} it was shown that different parts of the universe could bounce (or fail) in very different ways. It will be interesting to study such phenomenon with random perturbations.
\medskip\medskip\medskip

Although this paper considers both a contracting and expanding universe and how to connect them, the results are applicable to just $t \geq 0$:
\medskip\medskip

\noindent
{\it Purely Expanding Universe}
\medskip

\noindent It is noted that if we only consider a purely expanding universe for $t \geq 0$, without regard of how it got there, where the co-prime conditions are not considered and $w > 1$, then  the results of this paper provide a way to study the solutions of the Friedmann equations in the presence of random perturbations near the big bang. 
\medskip\medskip\medskip

\section{Conclusion} \label{sec:6}
\medskip

The stochastic regularization method developed  in this paper offers a new approach to studying the big bang singularity. It is interesting that the co-prime conditions on $w_c$ persist even in the presence of arbitrary random perturbation up to and including the big bang itself. This shows that these conditions are quite robust. Thus, although the co-prime conditions represent a fine tuning of the equation of state for solution extensions, their existence shows that they are structurally stable and hence may be physically significant. 

The results of this paper provide a rigorous demonstration of the extension of solutions through the big bang in the presence of random perturbations, which can model effects known and not known.

This paper was done for a particular model of the universe, however, the results may have a bearing on other models. It is interesting that near the big bang, the scale factor is not  smoothly  monotonic, but has random, jagged, non-differentiable variations  due to small stochastic perturbations. The stochastic modeling in this paper can also be applied a purely expanding universe.

It may be possible to experimentally detect the random, monotonic, non-smooth variations of $a(t)$. These variations may give rise to gravity waves, a topic of further study.

Although the relationship of the random perturbation model we have used here by a Brownian motion to quantum fluctuations was only briefly discussed, it would be interesting to further study this in more detail, which is out of the scope of this paper.

\ack

E.B. thanks David Spergel, Robert Vanderbei, and Paul Steinhardt for helpful discussions. B.X. is funded by Eric and Wendy Schmidt Foundation.

\appendix


\section*{References}

\begin{thebibliography}{99}

\bibitem{Xue:2014} 
  B.~K. ~Xue and E.~Belbruno,
   {\it Class.\ Quant.\ Grav.}\  {\bf 31}, 165002 (2014).
 
\bibitem{Mohammed:1999}
S-E.~Mohammed and M. ~Scheutzow,
{\it Ann.\ Probab.}\ {\bf 27}, 615 (1999).


\bibitem{Belbruno:2004}
  E.~Belbruno,
  {\it Capture Dynamics and Chaotic Motions in Celestial Mechanics},
  Princeton University Press, 2004.
  
  \bibitem{Garfinkle:2008ei} 
  D.~Garfinkle, W.~C.~Lim, F.~Pretorius and P.~J.~Steinhardt,
  {\it Phys.\ Rev.}\ D {\bf 78}, 083537 (2008).
  
    \bibitem{Erickson:2003zm} 
 J.~K.~Erickson, D.~H.~Wesley, P.~J.~Steinhardt and N.~Turok,
  {\it Phys.\ Rev.}\ D {\bf 69}, 063514 (2004).
  
\bibitem{Mukhanov:2005}
  V. ~Mukhanov,
  {\it Physical Foundations of Cosmology},
  Cambridge University Press, 2005. (p336)

\bibitem{Steinhardt:2001st} 
  P.~J.~Steinhardt and N.~Turok,
  {\it Phys.\ Rev.}\ D {\bf 65}, 126003 (2002).
  
  \bibitem{Lehners:2011kr} 
  J.~-L.~Lehners,
  {\it Class.\ Quant.\ Grav.}\  {\bf 28}, 204004 (2011).


\bibitem{McGehee:1981}
  R.~McGehee,
  {\it Comment.\ Math.\ Helveti.}\ {\bf 56}, 527-557 (1981).


\bibitem{Guckenheimer:2002}
  J.~Guckenheimer and P.~Holmes,
  {\it Nonlinear Oscillations, Dynamical Systems, and Bifurcations of Vector Fields},
  Springer-Verlag, 2002.

\bibitem{Belbruno:2013}
  E.~Belbruno,
  {\it Cel.\ Mech.\ Dyn.\ Astr.}\ {\bf 115}, 21-34 (2013).


\bibitem{Xue:2013}
  B.~Xue, D.~Garfinkle, F.~Pretorius, and P.~J.~Steinhardt,
  {\it Phys.\ Rev.}\ D {\bf88}, 083509 (2013).

\bibitem{Kloeden:1992}
  P.E. Kloeden and E. Platen,
  {\it Numerical Solution of Stochastic Differential Equations},
  Springer-Verlag, 1992.









































\end{thebibliography}
\end{document}